\documentclass[pre,twocolumn,a4paper,superscriptaddress,floatfix,10pt,nofootinbib,longbibliography]{revtex4-2}
\pdfoutput=1

\usepackage{amsfonts}
\usepackage{amssymb}
\usepackage{amsmath}
\usepackage{amsthm}
\usepackage{bm}
\usepackage{booktabs}
\usepackage{complexity}
\usepackage{dcolumn}
\usepackage[pdftex]{graphicx,color}
\usepackage{pdfsync}
\usepackage{rotate}
\usepackage{siunitx}
\usepackage{subfigure}
\usepackage{framed}
\usepackage[T1]{fontenc}

\usepackage[pdftex]{hyperref}

\begin{document}

\title{Boosting Monte Carlo simulations of spin glasses using autoregressive neural networks}

\author{B. McNaughton}
\affiliation{School of Science and Technology, Physics Division, Universit{\`a}  di Camerino, 62032 Camerino (MC), Italy}
\affiliation{Department of Physics, University of Antwerp, Groenenborgerlaan 171, B-2020 Antwerp, Belgium}

\author{M. V. Milo\v{s}evi{\'c}}
\affiliation{Department of Physics, University of Antwerp, Groenenborgerlaan 171, B-2020 Antwerp, Belgium}
\affiliation{NANOlab Center of Excellence, University of Antwerp, Belgium}

\author{A. Perali}
\affiliation{School of Pharmacy, Physics Unit, Universit{\`a} di Camerino, 62032 Camerino (MC), Italy}

\author{S. Pilati}
\affiliation{School of Science and Technology, Physics Division, Universit{\`a}  di Camerino, 62032 Camerino (MC), Italy}

\begin{abstract}
The autoregressive neural networks are emerging as a powerful computational tool to solve relevant problems in classical and quantum mechanics. 
One of their appealing functionalities is that, after they have learned a probability distribution from a dataset, they allow exact and efficient sampling of typical system configurations. 
Here we employ a neural autoregressive distribution estimator (NADE) to boost Markov chain Monte Carlo (MCMC) simulations of a paradigmatic classical model of spin-glass theory, namely the two-dimensional Edwards-Anderson Hamiltonian. 
We show that a NADE can be trained to accurately mimic the Boltzmann distribution using unsupervised learning from system configurations generated using standard MCMC algorithms. The trained NADE is then employed as smart proposal distribution for the Metropolis-Hastings algorithm. This allows us to perform efficient MCMC simulations, which provide unbiased results even if the expectation value corresponding to the probability distribution learned by the NADE is not exact.
Notably, we implement a sequential tempering procedure, whereby a NADE trained at a higher temperature is iteratively  employed as proposal distribution in a MCMC simulation run at a slightly lower temperature. This allows one to efficiently simulate the spin-glass model even in the low-temperature regime, avoiding the divergent correlation times that plague MCMC simulations driven by local-update algorithms.
Furthermore, we show that the NADE-driven simulations quickly sample ground-state configurations, paving the way to their future utilization to tackle  binary optimization problems.
\end{abstract}

\maketitle

\section{Introduction}
\label{intro}
Artificial neural networks are finding increasing applicability in various fields of classical and quantum physics~\cite{carleo2019machine}, where the generative neural networks are turning out to be particularly useful. They can be trained to mimic complex probability distributions, either using unsupervised learning protocols from unlabeled datasets, or using reinforcement learning schemes whereby a reward function is optimized.
Among other generative models (e.g., the variational autoencoders~\cite{rocchetto2018learning,luchnikov2019variational}), the restricted Boltzmann machines~\cite{ackley1985learning} were already proven successful in solving several computational tasks, including: learning classical thermodynamics from Monte Carlo samples~\cite{torlai2016learning}, accelerating classical Monte Carlo simulations~\cite{huang2017accelerated} (see also Ref.~\cite{liu2017self} for a related method), building accurate variational wave-functions~\cite{carleo2017solving}, performing quantum state tomography~\cite{torlaitomography}, simulating open quantum systems~\cite{nagy2019variational,PhysRevLett.122.250502,PhysRevLett.122.250503}, decoding topological codes~\cite{torlai2017neural}, guiding projective quantum Monte Carlo simulations~\cite{pilati2019self}, and reconstructing density matrices~\cite{carrasquilla2019reconstructing}.
The autoregressive neural networks provide additional distinctive functionalities compared to the restricted Boltzmann machines. 
In particular, owing to a specific connectivity structure, they allow writing the likelihood of any configuration as an ordered product of chained conditional distributions.
Therefore, by using ancestral sampling one can exactly sample system configurations according to the learned distribution. 
This avoids resorting to Markov chain Monte Carlo (MCMC) algorithms, which are often plagued by long correlations times, leading to an excessive computational cost in practical applications of generative sampling.
%
Ancestral sampling has already been exploited in quantum physics to accelerate the optimization of variational wave-functions~\cite{sharir2019deep,hibatallah2020recurrent}. Recently, a variational framework to solve rather general classical statistical-mechanics problems using autoregressive networks has also been presented~\cite{wu2019solving}. 
It has been applied to clean systems and also to a mean-field disordered model, namely the Sherrington-Kirkpatrick spin Hamiltonian~\cite{PhysRevLett.35.1792}. 
This model has infinite-range interactions, and its properties are exactly predicted by Parisi's mean-field theory based on the replica method~\cite{mezard1987spin}.
While the variational framework of Ref.~\cite{wu2019solving} extends well beyond the standard mean-field theories commonly employed to study spin glasses, it might still provide biased results, since the probability distribution learned by the neural network does not necessarily exactly coincide with the Boltzmann distribution.

This bias can be eliminated with two approaches. In the first, the autoregressive model is used for importance sampling in a re-weighting scheme. In the second, it is used as a smart proposal distribution for the Metropolis-Hastings algorithm. 
In the field of machine learning, the first approach has been employed to compute otherwise intractable normalization integrals~\cite{papamakarios2015distilling}. An application of both approaches to classical statistical mechanics has appeared only very recently~\cite{nicoli2019asymptotically}\footnote{A generative neural network has been used to propose updates in a MCMC simulation also in Ref.~\cite{huang2017accelerated}, but using a restricted Boltzmann machine instead of an autoregressive model. This required to run a parallel MCMC simulation, which was implemented via alternated Gibbs sampling.}.
However, the study of Ref.~\cite{nicoli2019asymptotically}  focused only on clean ferromagnetic models, and it considered only the reinforcement learning scheme to train the neural network, as in Ref.~\cite{wu2019solving}.

Random spin systems with frustrated interactions display many intriguing phenomena related to glass physics~\cite{binder1986spin,chowdhury2014spin}, including:  magnetic correlations, replica symmetry breaking, hysteresis, and aging.
In fact, spin glass models with short-range interactions  have challenged theoretical physicists for decades. 
It is unclear whether the mean-field replica theory, which exactly describes infinite-range models, applies also to short-range systems, even at the qualitative level (see, e.g., Refs.~\cite{marinari2000replica,PhysRevE.61.1121,wang2017number,PhysRevB.90.184412}).
The computational problems originate from the exceedingly long autocorrelation times that plague standard MCMC simulations driven by local-update algorithms. 
This problem arises also when addressing binary optimization problems --- which are ubiquitous in scientific research and in industry --- via stochastic optimization methods such as simulated annealing~\cite{kirkpatrick1983optimization}. 
Indeed, identifying the optimal solution is equivalent to finding the lowest-energy configuration of a disordered Ising Hamiltonian. 
%
%
In the last decades, relevant algorithmic developments have occurred in the field of spin glasses. In particular, one should mention the global-update methods such as the parallel tempering technique~\cite{hukushima1996exchange} and the isoenergetic cluster updates~\cite{houdayer2001cluster,zhu2015efficient}. Still, novel and possibly more flexible MCMC methods would be extremely useful.

In this article we investigate the use of autoregressive neural networks to increase the efficiency of MCMC simulations of spin glasses. 
The model we focus on is a paradigmatic short-range spin model, namely the two-dimensional Edwards-Anderson Hamiltonian. The nearest-neighbor couplings are sampled from a gaussian distribution. 
%
The neural network we employ is a standard autoregressive generative model, namely the so-called neural autoregressive distribution estimator (NADE)~\cite{larochelle2011neural}. 
In this article, the network is trained in an unsupervised learning scheme, which consists of minimizing the Kullback-Leibler divergence with respect to a set of spin configurations sampled using MCMC simulations driven by a standard local-update algorithm.
Our analysis shows that the NADE can learn to accurately mimic the Boltzmann distribution of the spin-glass model. We quantify this accuracy, and how it varies with the number of hidden neurons and the physical system size, by comparing the energy expectation-value corresponding to the NADE with the exact result corresponding to the Boltzmann distribution.
The trained NADE is then employed as proposal distribution for the Metropolis-Hastings algorithm. This allows us to essentially eliminate the correlations that affect standard MCMC simulations driven by local-update algorithms.
An important contribution of this article is the implementation of a sequential tempering procedure. This starts from a moderately high temperature, where uncorrelated configurations are easily sampled also via the local-update MCMC algorithm. Then, it performs a sequence of MCMC simulations at successively lower temperatures, whereby the NADE trained at the previous temperature is used to drive the Metropolis-Hastings algorithm. 
This allows us to perform efficient MCMC simulations even in the low temperature regime, where local-update algorithms become impractical due to the diverging autocorrelation times.
Finally, we analyze how efficiently the NADE-driven simulations performed at low temperatures sample the ground-state configurations. The obtained encouraging results lead us to advocate the use of autoregressive models to boost stochastic optimization methods such as the simulated annealing.

The article is organized as follows: in Section~\ref{method} we describe the Ising glass Hamiltonian, the NADE training method, the local-update as well as the NADE-driven MCMC algorithms. Section~\ref{results} presents the results obtained by training the NADE, by running local-update and NADE-driven MCMC simulations, and by performing the sequential tempering procedure. The possible use of NADEs to boost stochastic optimization algorithms is also discussed.
Our conclusions and some future perspectives are given in Section~\ref{conclusions}.

\begin{figure}[t]
\begin{center}
\includegraphics[width=\linewidth]{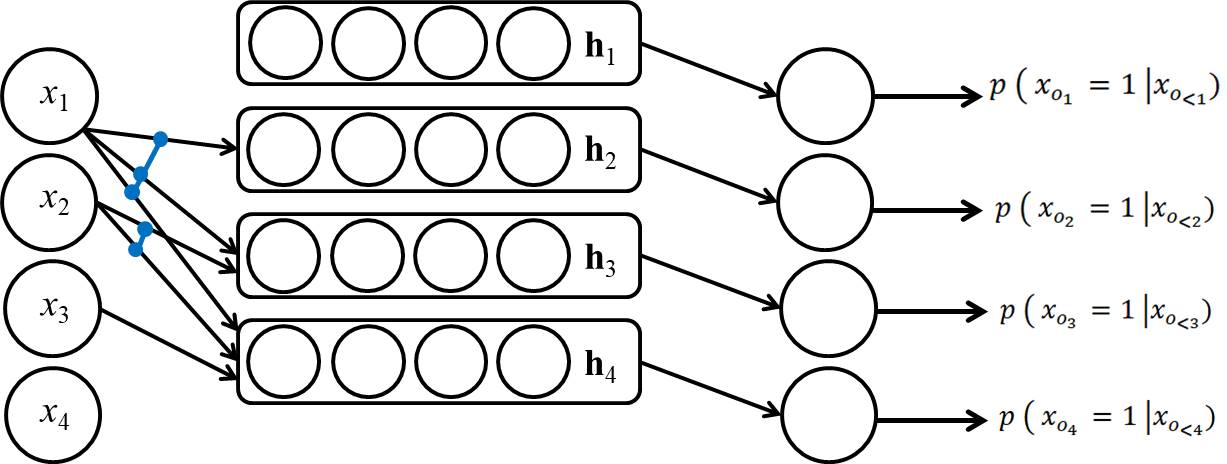}
\caption{
Illustration of a neural autoregressive
distribution estimator. Arrows connected by 
blue segments correspond to connections with shared 
parameters.
}
\label{fig:NADE}
\end{center}
\end{figure}

\section{Model and methods}
\label{method}
The spin-glass model addressed in this article is the two-dimensional Edwards-Anderson Hamiltonian:
\begin{equation}
\label{H}
H( \boldsymbol{ \sigma}) = - \sum_{\left<ij\right>}J_{ij}\sigma_{i}\sigma_{j},
\end{equation}
where $\sigma_{i}\in\{-1,1\}$ are binary spin variables at the sites labeled by the indices $i,j=1,\dots,N$, $\boldsymbol{\sigma}=(\sigma_1,\dots,\sigma_N)$ indicates the spin configuration, and $J_{ij}$ is the coupling strength between the spins $i$ and $j$. We consider random couplings sampled from a gaussian distribution with zero mean and unit variance. The sum in the above equation runs over nearest-neighbor sites on a square lattice. Periodic boundary conditions are adopted.
The thermodynamic properties, e.g., the average energy $E$, are computed as $E=\langle H( \boldsymbol{\sigma})\rangle$, where the angular brackets indicate expectation values over the Boltzmann distribution
$
P\left ( \boldsymbol{\sigma } \right ) = \exp\left[ -\beta H(\boldsymbol{\sigma}) \right]/Z.
$
Here, $\beta = 1/k_{B}T$ is the inverse temperature, $k_B=1$ is the Boltzmann constant, $T$ is the temperature, and $Z = \sum_{\boldsymbol{\sigma }} \exp\left[-\beta H(\boldsymbol{\sigma })\right]$ is the partition function.
Expectation values of this kind can be computed by implementing a stochastic Markov chain in the configuration space driven by a transition matrix. The entries of this matrix will be denoted as $T_{\boldsymbol{\sigma}\prime \boldsymbol{\sigma}}$. They must satisfy the conditions
$T_{\boldsymbol{\sigma}\prime \boldsymbol{\sigma}} \geqslant 0$ and
 $\sum_{\boldsymbol{\sigma}\prime}T_{\boldsymbol{\sigma}\prime \boldsymbol{\sigma}}=1$, for any $\boldsymbol{\sigma}$, meaning that the transition matrix is stochastic. 
$T_{\boldsymbol{\sigma}\prime \boldsymbol{\sigma}}$ represents the probability to move to the configuration $\boldsymbol{\sigma}\prime$ from $\boldsymbol{\sigma}$.
A common procedure is to decompose the transition matrix in the proposal-acceptance form: $T_{\boldsymbol{\sigma}\prime \boldsymbol{\sigma}}=\omega_{\boldsymbol{\sigma}\prime \boldsymbol{\sigma}}A_{\boldsymbol{\sigma}\prime \boldsymbol{\sigma}}$. The entries of the proposal distribution matrix  $\omega_{\boldsymbol{\sigma}\prime \boldsymbol{\sigma}}$ represent the probability to propose moving to $\boldsymbol{\sigma}\prime$ from $\boldsymbol{\sigma}$. This matrix must be stochastic, and satisfy an ergodic condition, meaning that it is possible to reach any configuration  from any other in a finite number of steps. $A_{\boldsymbol{\sigma}\prime \boldsymbol{\sigma}}$ is the probability to accept the proposed update; the probability to reject it, thus iterating the old configuration $\boldsymbol{\sigma}$ in the Markov chain, is $1-A_{\boldsymbol{\sigma}\prime \boldsymbol{\sigma}}$.
One way to satisfy the detailed balance condition --- which is sufficient (although not necessary) to ensure that the Boltzmann distribution is the stationary distribution of the Markov chain --- is to define the acceptance probability as:
\begin{equation}
\label{MH}
A_{\boldsymbol{\sigma}\prime \boldsymbol{\sigma}} = \min \left( 1, \frac{P\left ( \boldsymbol{\sigma }\prime \right ) \omega_{\boldsymbol{\sigma} \boldsymbol{\sigma}\prime}} {P\left ( \boldsymbol{\sigma } \right ) \omega_{\boldsymbol{\sigma}\prime \boldsymbol{\sigma}} } \right).
\end{equation}
This formula corresponds to the famous Metropolis-Hastings algorithm~\cite{10.1093/biomet/57.1.97}. 
Notice that, since one needs only ratios of Boltzmann-distribution values, the normalization factor $Z$ is not needed.
A common choice is to consider a symmetric proposal distribution, i.e. $\omega_{\boldsymbol{\sigma} \boldsymbol{\sigma}\prime} = \omega_{\boldsymbol{\sigma} \prime\boldsymbol{\sigma}}$. In this case the acceptance probability simplifies to: $A_{\boldsymbol{\sigma}\prime \boldsymbol{\sigma}} = \min \left( 1, \frac{P\left ( \boldsymbol{\sigma }\prime \right ) } {P\left ( \boldsymbol{\sigma } \right )  } \right)$. 
For example, one can randomly choose a single spin $i$ and propose to flip it, setting ${\sigma\prime}_i=-\sigma_i$. This corresponds to the matrix entries: $\omega_{\boldsymbol{\sigma} \prime\boldsymbol{\sigma}}=1/N$ if $\boldsymbol{\sigma}$ and $\boldsymbol{\sigma}\prime$ differ by one spin-flip only, and $\omega_{\boldsymbol{\sigma} \prime\boldsymbol{\sigma}}=0$ otherwise.
In what follows, this local method will be referred to as single spin-flip algorithm. It is efficient enough for rather generic models. However, in the vicinity of phase transitions (e.g., ferromagnetic transitions in ordered Ising models) or in the glassy phases of disordered systems, the dynamics of MCMC simulations driven by the single spin-flip  algorithm suffer a pathological slowing down, possibly leading to the breakdown of ergodicity.
This slowing down is associated to strong statistical correlations between configurations  subsequently sampled along the Markov chain.
In particular, in the case of low-temperature spin-glass models, the correlation time diverges and the Markov chain is not ergodic, meaning that not all physically relevant regions of the configuration space are explored in the feasible computational times.
For certain relevant systems, these correlations can be suppressed adopting more sophisticated (in general, non-symmetric) proposal distributions $\omega_{\boldsymbol{\sigma} \prime\boldsymbol{\sigma}}$. This approach is often referred to as smart Monte Carlo method~\cite{thijssen2007computational}.
For example, for ferromagnetic Ising models one can adopt the Swendsen-Wang or the Wolff algorithms~\cite{PhysRevLett.58.86,PhysRevLett.62.361}. These perform cluster moves instead of single spin-flip updates. The worm algorithm is a relevant alternative~\cite{prokof2001worm,deng2007dynamic}.
MCMC methods that perform significantly better than the single spin-flip updates have been developed also for spin glasses. Relevant examples are the parallel tempering method~\cite{hukushima1996exchange} and the iso-energetic cluster-update algorithms~\cite{houdayer2001cluster,zhu2015efficient}. 
However, spin glasses still represent a computational challenge. 
The long correlation times plague also most heuristic methods commonly employed to solve binary optimization problems. In fact, identifying the optimal solution is equivalent to finding one of the ground-state configurations of a spin-glass model. 
This task constitutes a non-deterministic polynomial hard problem when implemented on a non-planar graph~\cite{barahona1982computational}.
The archetypal heuristic optimization method is simulated annealing~\cite{kirkpatrick1983optimization}.
This method exploits MCMC algorithms to explore possible solutions, but the lack of ergodicity might prevent the dynamics from reaching the lowest-energy configuration.
Below we report how to exploit autoregressive neural networks to implement efficient MCMC algorithms for Ising Hamiltonians of the type defined in equation~\eqref{H}. This the central objective of this work.

\subsection{The neural autoregressive distribution estimator (NADE)}
An efficient proposal distribution can be constructed using autoregressive neural networks. 
In this article, we consider the use of NADEs. 
Like other generative neural networks, NADEs can be trained to model complex probability distribution from sampled data. Then, they allow direct sampling of system instances from the learned probability distribution. 
The system instances are represented by vectors of binary variables $\boldsymbol{x}=(x_1,\dots,x_D)$. Here, following the convention of the machine-learning literature, we consider the binary values $x_{d} \in \left \{ 0,1 \right \}$, for $d=1,\dots,D$.
The joint distribution of the binary variables is decomposed as a product of chained conditional probabilities:
\begin{equation}
\label{prob}
p(\bold{x}) = \prod_{d=1}^{D}p(x_{o_{d}}|\bold{x}_{o_{<d}}).
\end{equation}
In the above equation, $o$ denotes the chosen ordering of the binary variables, $o_{d}$ indicates the $d$-th variable in the ordering $o$, and the slice subscript $o_{<d}$ indicates the first $d-1$ dimensions in $o$; therefore, $\bold{x}_{o_{<d}}$ represents the sub-vector including the indicated dimensions only.
Each conditional probability is modeled using a feed-forward neural network, defined as:
\begin{equation}
p(x_{o_{d}}=1|\bold{x}_{o_{<d}}) = \mathrm{sigm}(\bold{V}_{o_{d},\cdot}\bold{h}_{d}+b_{o_{d}}),
\end{equation}
where $\bold{V}$ is a weight matrix, $b_{o_{d}}$ is the bias, and one has $D$ vectors of hidden-neuron activations, computed as
\begin{equation}
\bold{h}_{d} = \mathrm{sigm}(\bold{W}_{\cdot ,o_{<d}}\bold{x}_{o_{<d}}+\bold{c}),
\end{equation}
with the weight matrix $\bold{W}$ and the bias vector $\bold{c}$. The activations are computed with the logistic function $\mathrm{sigm}(x)= 1/\left[1+\exp(-x)\right]$. We used the notation of vectorized functions and that of slice indices also for sub-matrices. 
The NADE has $D$ hidden layers, each including $N_H$ neurons. An important property is that the matrix $\bold{W}$ and the bias vector $\bold{c}$ are shared by all hidden layers. 
The parameters to be optimized in the training process are $\bold{V} \in \mathbb{R}^{D\times N_H}$, $\bold{b} \in \mathbb{R}^{D}$, $\bold{W} \in \mathbb{R}^{N_H\times D}$, and $\bold{c} \in \mathbb{R}^{N_H}$. 
The structure of the whole neural network is represented in Fig.~\ref{fig:NADE}.
By construction, the conditional probability distribution $p(x_{o_{d}}|\bold{x}_{o_{<d}})$ for the variable $x_{o_d}$ does not depend on the subsequent variables $\bold{x}_{o_{>d}}$ in the ordering $o$.
Therefore, after training has been performed, one can sample configurations $\bold{x}$ from the learned distribution $p(\bold{x})$ via ancestral sampling: following the ordering $o$, each variable $x_{o_d}$ is sampled from the binary distribution $p(x_{o_{d}}|\bold{x}_{o_{<d}})$, computed using the previously sampled variables.
This would not be possible with a closely related neural-network model such as the restricted Boltzmann machine. In fact, in that case one has to resort to MCMC algorithms, usually implemented via alternated Gibbs sampling of hidden neurons and visible neurons~\cite{hinton2002training}. This might cause problems associated to long correlation times, leading to an excessive computational cost in practical application of generative sampling.
Furthermore, with the NADE the (normalized) likelihood of a configuration $\bold{x}$ can be efficiently computed via the formula~(\ref{prob}). Instead, with the restricted Boltzmann machine one has to determine the normalization integral, namely the partition function, which is an intractable computation already for moderately large systems.
The NADE can be trained in an unsupervised learning scheme from a (typically large) dataset $\{\bold{x}^{(n)}\}$, where $n=1,\dots,N_t$, and $N_t$ is the training-dataset size. The cost function to be minimized is the average negative log-likelihood, given by
\begin{equation}
\mathrm{nl}(\bold{V},\bold{b},\bold{W},\bold{c})= -\frac{1}{N_t}\sum_{n=1}^{N_t} \log p(\bold{x}^{(n)}).
\end{equation}
It can be shown that this criterion corresponds to the minimization of the  so-called 
Kullback-Leibler divergence (see, e.g.,~\cite{fischer2012introduction}), which is defined as:
\begin{equation}
\mathrm{KL} \left( q \left| \right| p\right) = \sum_{\bold{x}} q(\bold{x}) \ln\left(q(\bold{x}) /p(\bold{x}) \right),
\end{equation}
where $q(\bold{x})$ is the underlying probability distribution of the samples $\{\bold{x}^{(n)}\}$, which is in general unknown.
The optimization of the network parameters $\boldsymbol{\Theta}\equiv \{\bold{V},\bold{b},\bold{W},\bold{c}\}$ is performed via the stochastic gradient descent algorithm. 
Starting from reasonably chosen initial values $\boldsymbol{\Theta}_0$, at each step $s=0,1,\dots$ one performs the update $\boldsymbol{\Theta}_{s+1}=\boldsymbol{\Theta}_{s}-\eta \nabla_{\boldsymbol{\Theta}}  \mathrm{nl}(\boldsymbol{\Theta}){|_{\boldsymbol{\Theta}_{s}}} $, where the scalar $\eta$ is the learning rate. The log-likelihood gradient is computed via the back-propagation algorithm. In the standard implementation, at each step only a small mini-batch of $N_b$ system instances, randomly sampled from the training set, is used to compute the gradient. One learning epoch includes  $\lfloor N_t/N_b \rfloor$ steps ($\lfloor \;\;\rfloor$ is the floor function), and the optimization is iterated for a number of epochs $N_E$, till convergence is reached. 
In our implementation, a large training set is considered, and in each epoch the mini-batches are sampled from a random subset of the total training set, as detailed in the next section.
For the specific problems addressed in this article, it appears that regularization procedures are not strictly necessary.
All details and the pseudocode of the algorithm to compute the log-likelihood and its gradient can be found in Refs.~\cite{larochelle2011neural,uria2016neural}.
It is worth mentioning that this algorithm exploits the NADE's specific structure, in particular the sharing of $\bold{W}$ and $\bold{c}$, to enhance efficiency.
Specifically, the computational cost to compute the likelihood of a system instance scales as $N_H D$, instead of the $N_H D^2$ scaling corresponding to the na\"ive implementation that does not exploit the NADE's structure~\cite{larochelle2011neural,uria2016neural}.

In this article, a NADE is used to learn the probability distribution corresponding to a large dataset of equilibrium configurations of the spin-glass model (\ref{H}). These configurations are generated using a single spin-flip MCMC simulation. In regimes where this local algorithm performs an ergodic dynamics, this simple procedure is sufficient to allow the NADE to learn the Boltzmann distribution.
To reach also the regimes where the dynamics given by the local algorithm are not ergodic, we implement a sequential tempering procedure. This is described in the next section.
For our purpose, the size of the network input-layer has to be $D=N$. The spins are ordered line-by-line. To switch between the two binary-value conventions, namely $\sigma_i\in\{-1,1\}$ and $x_i \in\{0,1\}$, the following mapping is used: $x_i=1$ if $\sigma_i=1$, and $x_i=0$ if $\sigma_i=-1$.
Once the training has been performed, the NADE probability distribution $p(\boldsymbol{\sigma})$ approximates the Boltzmann distribution $P(\boldsymbol{\sigma})$. 
Therefore, the thermodynamic properties as, e.g., the average energy $E$, can be approximated by expectation values over the NADE probability distribution: 
\begin{equation}
\label{Enade}
E_{\mathrm{NADE}} \simeq \langle H(\boldsymbol{\sigma})\rangle_{p(\boldsymbol{\sigma})} = \lim_{N_s\rightarrow\infty} \frac{1}{N_s} \sum_{n=1}^{N_s} H(\boldsymbol{\sigma}^{(n)}),
\end{equation}
where the $N_s$ configurations $\boldsymbol{\sigma}^{(n)} \sim p(\boldsymbol{\sigma}^{(n)})$ are efficiently sampled from the NADE distribution using ancestral sampling. In practice, one uses a large but finite number of configurations $N_s$. This procedure is expected to be very efficient, since the sampled configurations are perfectly uncorrelated.
As shown in the next section, this calculation allows us to quantify how accurately the NADE approximates the Boltzmann distribution. 
In general, a NADE with a finite number of hidden neurons $N_H$ will not exactly match the Boltzmann distribution, so the expectation values over the NADE distribution will be biased. Note that, even if $N_H$ is very large, a bias might also originate from an imperfect training due to a too small training set or to a failure of the optimization algorithm.
Two strategies can be adopted  to remove this bias.
The first one, which we employ in this article, is to combine the trained NADE with a standard MCMC method, as anticipated above. Specifically, we use the NADE as proposal distribution for the Metropolis-Hastings algorithm; that is, we set $\omega_{\boldsymbol{\sigma}\prime\boldsymbol{\sigma}}=p(\boldsymbol{\sigma}\prime)$. 
This procedure is correct if $p(\boldsymbol{\sigma})$ and $P(\boldsymbol{\sigma})$ have the same support. Formally, this condition is always satisfied since the conditional probabilities are computed using the logistic function, which has values $\sigma(x)\in(0,1)$ for any finite $x$. In the next section we analyze if and when the weight of the NADE distribution is sufficiently large, in any physically relevant configuration, to produce an efficient simulation.
It is worth stressing that, with this choice, the proposal distribution $\omega_{\boldsymbol{\sigma}\prime\boldsymbol{\sigma}}$ is independent of the starting configuration $\boldsymbol{\sigma}$. Therefore, when the proposal is accepted, the next configuration in the Markov process is  uncorrelated with the previous one. If 
$p(\boldsymbol{\sigma})=P(\boldsymbol{\sigma})$ the acceptance probability is $A_{\boldsymbol{\sigma}\prime\boldsymbol{\sigma}}=1$. Therefore, statistical correlations along the Markov chain only originate from the approximation in the NADE distribution, since this leads to some rejections. Our results, shown in the next section, indicate that NADEs can be  accurate enough to have a high acceptance rate and, therefore, enable efficient MCMC simulations of the spin-glass model.
One should also note that, since the computational cost of computing the configuration likelihood and the conditional probabilities (required for ancestral sampling) is linear in the system size, the NADE allows one to implement an efficient global-update algorithm with a linear computational cost.

The second strategy to remove the bias in the NADE expectation values is based on a re-weighting scheme. In fact, an unbiased expectation value can be computed as~\cite{nicoli2019asymptotically} 
\begin{equation}
E= \lim_{N_s\rightarrow\infty} \sum_{n=1}^{N_s} z_nH(\boldsymbol{\sigma}^{(n)}) ,
\end{equation}
where $\boldsymbol{\sigma}^{(n)} \sim p(\boldsymbol{\sigma}^{(n)})$, the weights are $z_n=\zeta_n/\sum_{n=1}^{N_s} \zeta_n$, with
$\zeta_n=\exp[-H(\boldsymbol{\sigma}^{(n)})]/p(\boldsymbol{\sigma}^{(n)})$. 
An analogous re-weighting strategy has been used in Ref.~\cite{papamakarios2015distilling} to compute the partition function of a restricted Boltzmann machine. This computation would otherwise be intractable.
Very recently, in Ref.~\cite{nicoli2019asymptotically} both strategies described above have been adopted, but focusing only on the thermodynamic properties of a clean ferromagnetic Ising model. 
Interestingly, Ref.~\cite{nicoli2019asymptotically} discusses also how to determine expectation values that explicitly depend on the partition function, as, e.g, the entropy, including the correct formula for the corresponding statistical uncertainty. In the study of Ref.~\cite{nicoli2019asymptotically}, the neural network has been trained via a reinforcement learning procedure, as in Ref.~\cite{wu2019solving}. This reinforcement procedure consists of minimizing a variational ansatz for the free energy based on the generative neural network. The training batches are sampled from the neural network distribution, rather than from a MCMC simulation. 
Rather than a NADE, Ref.~\cite{nicoli2019asymptotically} employed an autoregressive neural network named PixelCNN~\cite{oord2016pixel}, as in Ref.~\cite{wu2019solving}.

\begin{figure}[t]
\begin{center}
\includegraphics[width=\linewidth]{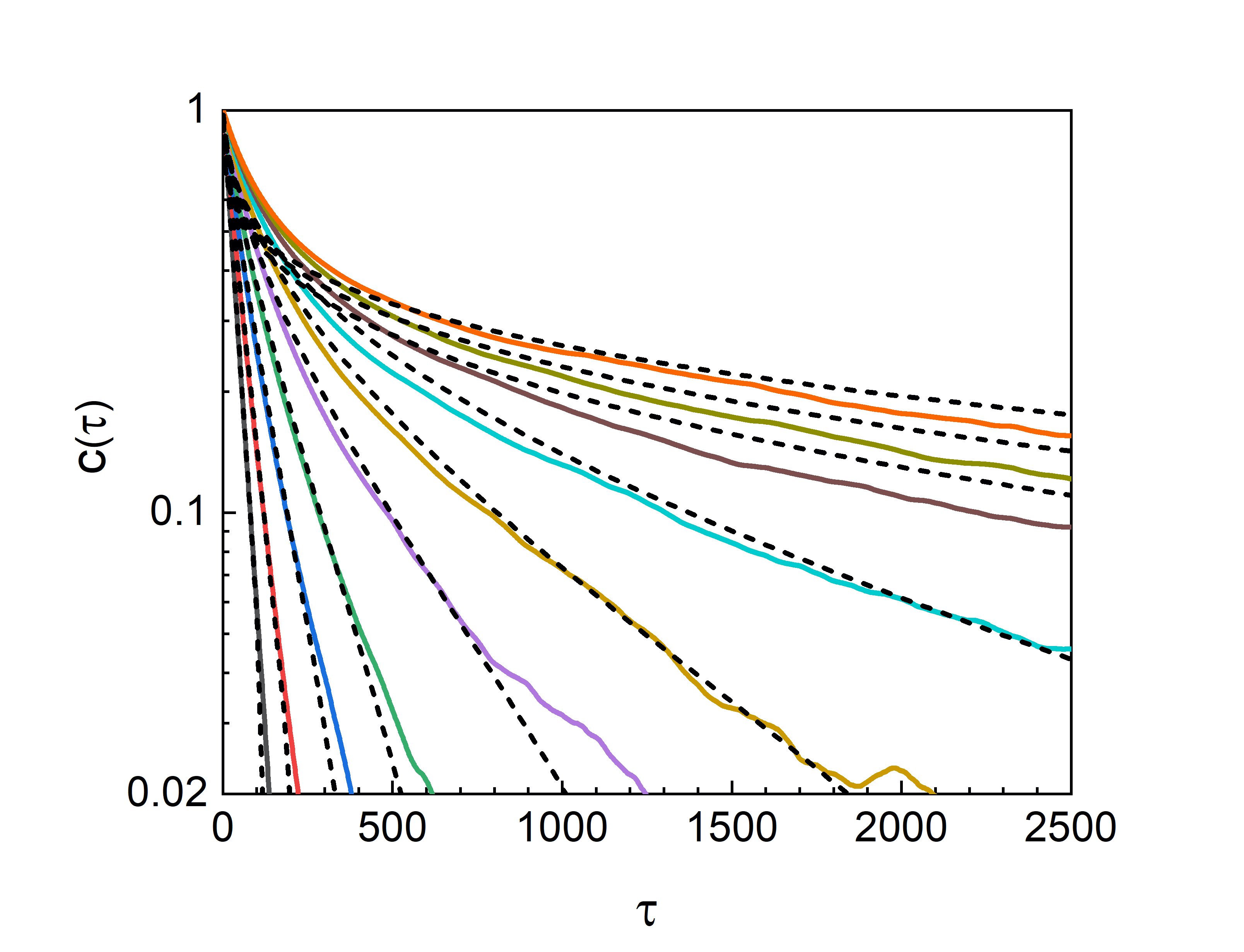}
\caption{Autocorrelation function $c(\tau)$ of the configuration energies $H$ obtained from single spin-flip MCMC simulations at temperatures $\beta=0.1,0.1,\dots,1$ (bottom to top). Dashed lines represent stretched-exponential fitting functions.
}
\label{fig:corrfun}
\end{center}
\end{figure}

\section{Results}
\label{results}
The spin-glass model defined in Eq.~\eqref{H} can be simulated using the single spin-flip Metropolis-Hastings algorithm, as explained in the previous section.
However, it is well known that such simulations can be affected by long autocorrelation times along the Markov chain~\cite{binder1986spin}, in particular in the low-temperature regime. 
To illustrate this effect and to quantify these statistical correlations, we compute the autocorrelation function $c(\tau)$ of the configuration energy. $c(\tau)$ is defined as:
\begin{equation}
c(\tau) = \frac{\langle H(t+\tau) H(t)\rangle - \langle H(t) \rangle^2}{\langle H(t)H(t)\rangle - \langle H(t) \rangle^2},
\end{equation}
where integers $\tau$ and $t$ count MCMC steps, $H(t)$ is the energy in the spin configuration sampled at the $t$-th step of the Markov chain, and the angular brackets indicate the average over the whole Markov chain. 
Following the standard procedure, we disregard an initial segment to account for equilibration effects.
The results corresponding to a typical realization of the gaussian random couplings are shown in Fig.~\ref{fig:corrfun}. Different realizations provide qualitatively similar results. The correlation function appears to be reasonably well described by an empirical stretched-exponential fitting function: $c(\tau) = a\exp[-(\tau/\tau^*)^\alpha]$. Here, $a$, $\tau^*$, and $\alpha$ are fitting parameters. At relatively low inverse temperatures $\beta \lesssim 0.5$, we obtain $\alpha\simeq 1$, corresponding to the common exponential decay, and $\tau^* \approx 50$, indicating that the correlation time is sufficiently short. Indeed, MCMC simulations much longer than $\tau^*=50$ are feasible, even with modest computational resources.
However, already at $\beta\simeq 1$, we obtain $\alpha \simeq 0.3 $ and $\tau^* \simeq 400$. For inverse temperatures $\beta > 1$, computing the thermodynamic properties with the single spin-flip updates becomes hard or absolutely unfeasible  due to the breakdown of ergodicity for the feasible simulation times in the low-temperature regime. 
%
%
In the early literature on spin-glasses it had been conjectured that this freezing temperature  $\beta \simeq 1$ was associated to a spin-glass transition (see, e.g., Ref.~\cite{binder1976phase}, and also Ref.~\cite{binder1986spin} for a review). However, later studies based on faster computers and on more efficient algorithms have clarified that for two-dimensional Ising Hamiltonian with nearest-neighbor interactions, a proper spin-glass phase --- identified, among other criteria, by the Edwards-Anderson order parameters or from the distribution of overlaps among system replicas --- occurs only in the zero-temperature limit~\cite{rieger1996critical,kawashima1997finite,hartmann2001lower,houdayer2001cluster,carter2002aspect}.

\begin{figure}[b]
\begin{center}
\includegraphics[width=\linewidth]{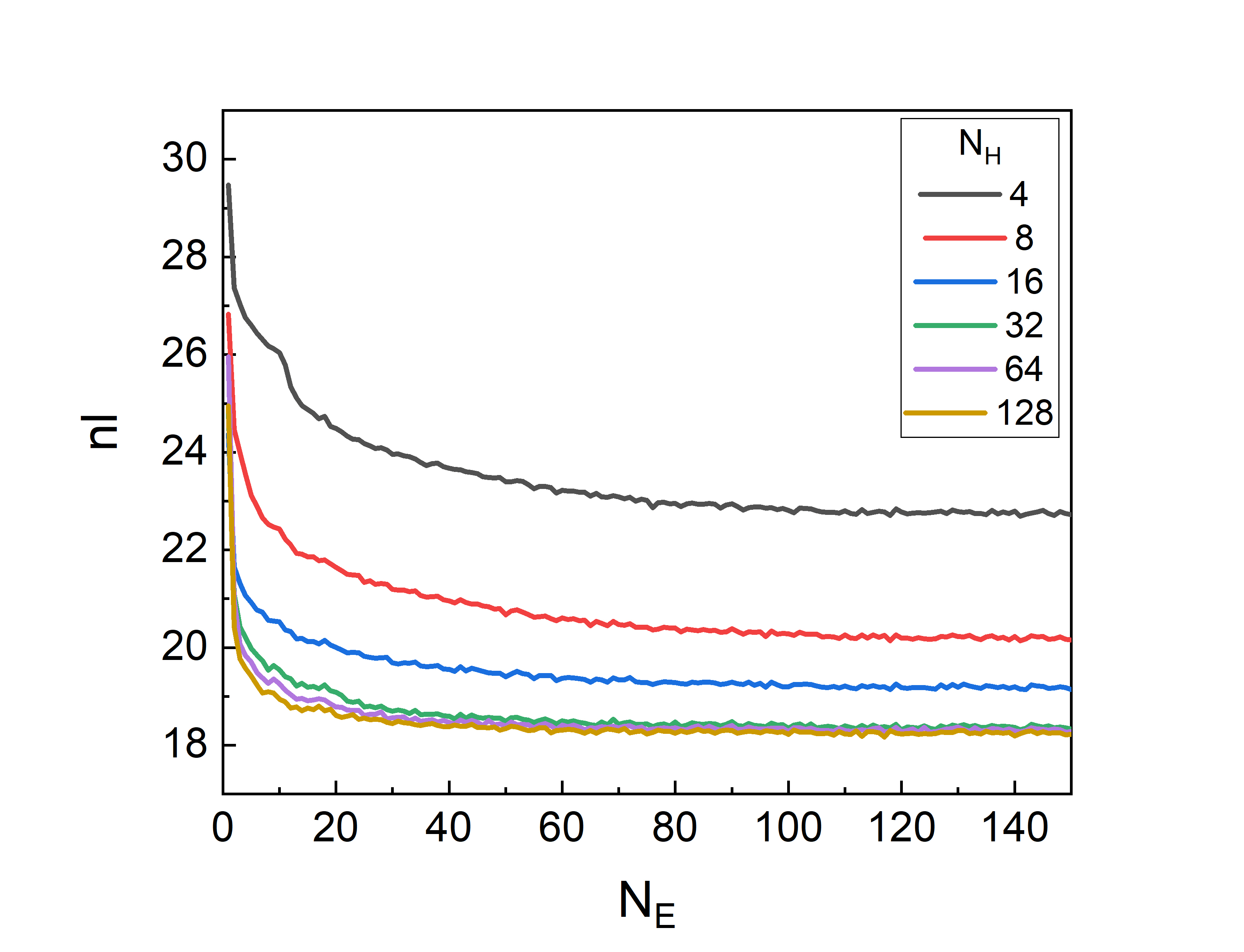}
\caption{Average of the negative log-likelihood ($\mathrm{nl}$) as a function of the number of epochs ($N_E$). Different curves correspond to different numbers of hidden units $N_H$, increasing from top to bottom. The size of the two-dimensional spin glass is $N=100$ and the inverse temperature is $\beta=1$.
}
\label{fig:loglikelihoodH}
\end{center}
\end{figure}

\begin{figure}[t]
\begin{center}
\includegraphics[width=\linewidth]{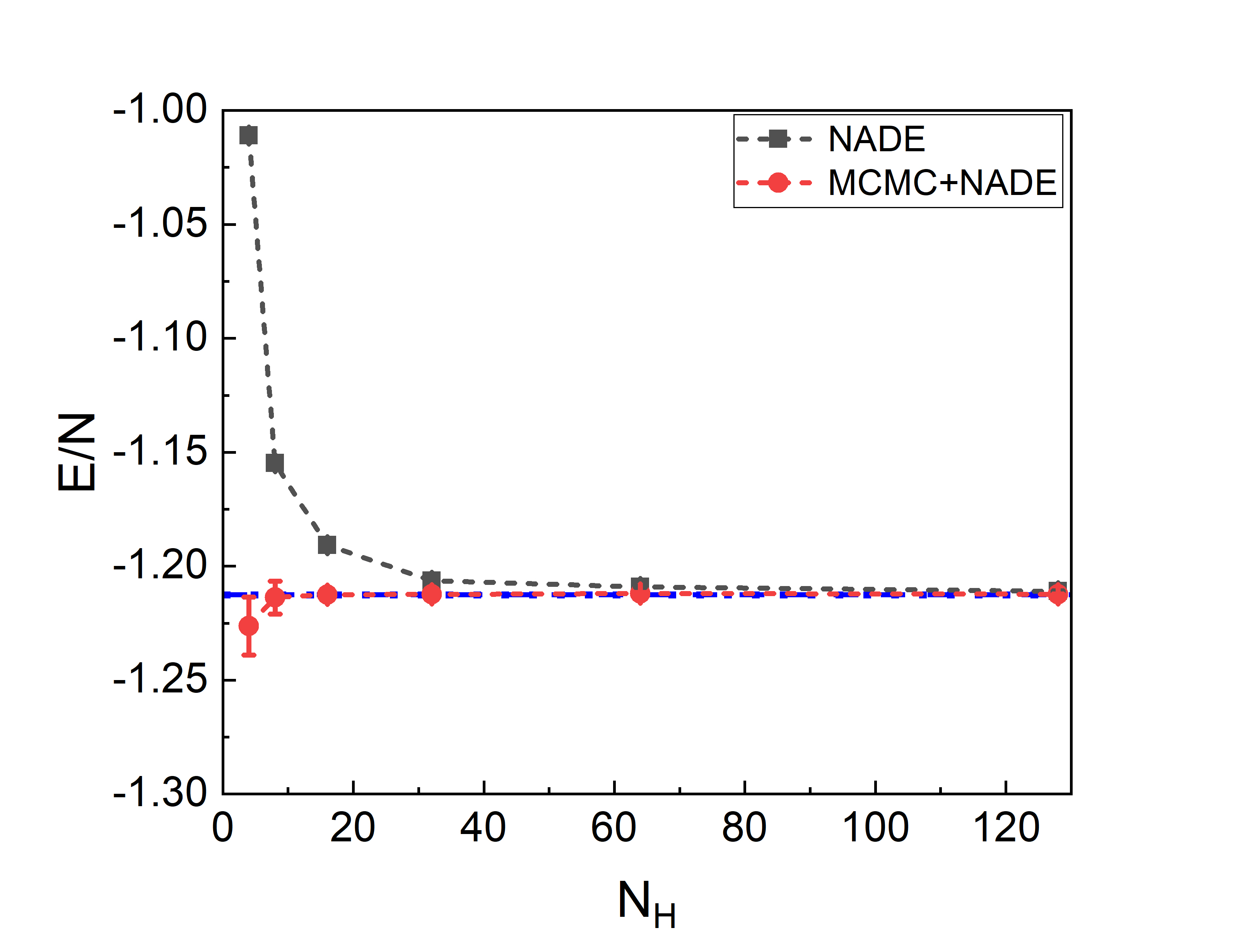}
\caption{Average energy per spin $E/N$ as a function of the number of hidden units $N_H$. The system size is $N=100$ and the inverse temperature is $\beta=1$.
The black squares correspond to the expectation value over the probability distribution learned by the NADE (see Eq.~\eqref{Enade}). The red dots correspond to the results of the NADE-driven MCMC simulation. The horizontal blue line indicates the exact result.
}
\label{fig:Enade}
\end{center}
\end{figure}

\begin{figure}[t]
\begin{center}
\includegraphics[width=\linewidth]{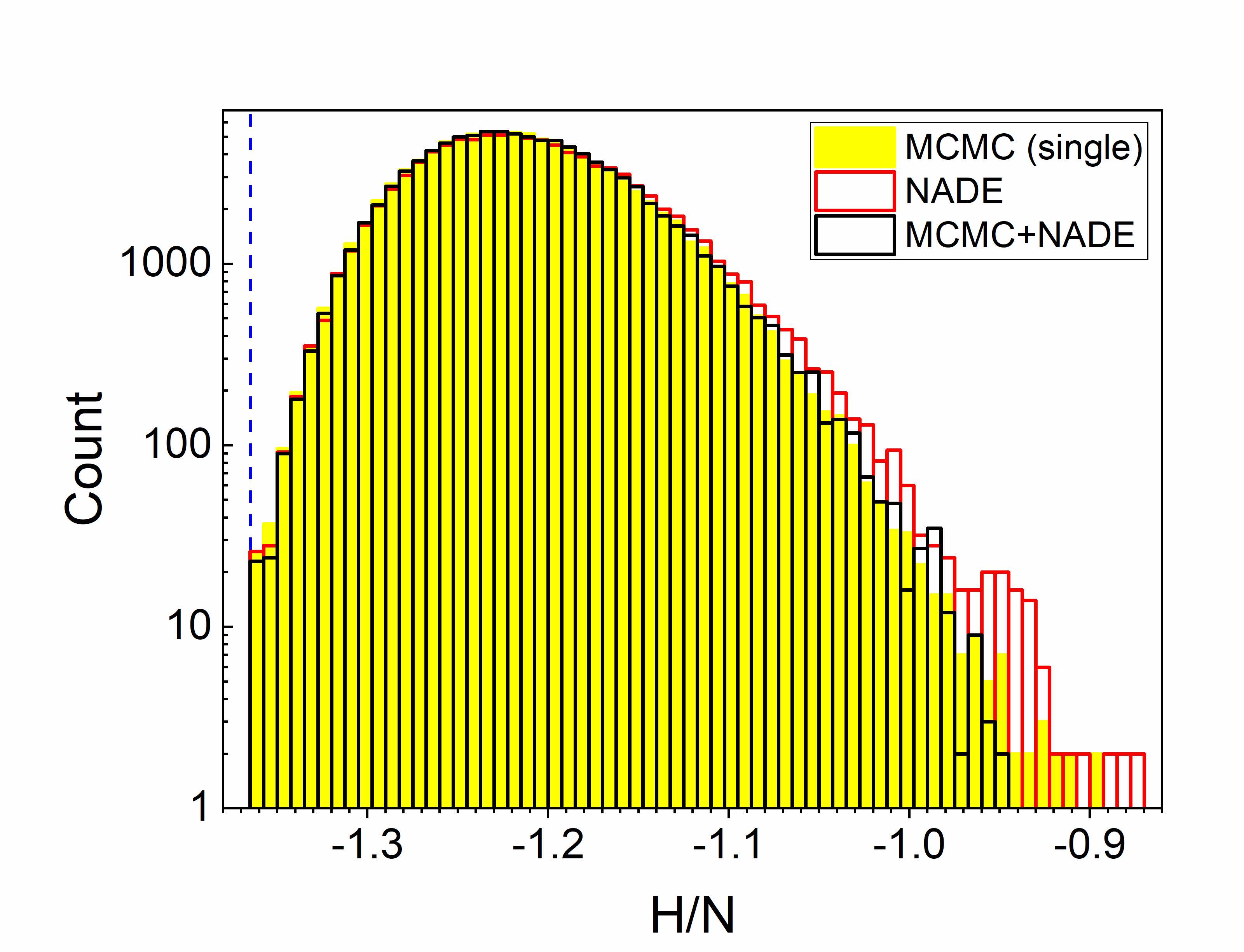}
\caption{Histogram of $10^5$ sampled configuration energies per spin $H/N$. The full yellow columns correspond to a single spin-flip MCMC simulation ($8\times 10^7$ steps, every $800$-th configuration is counted), the empty black columns to a NADE-driven MCMC simulation (every $10$-th configuration is counted), and the empty red columns to the energies sampled  via ancestral sampling from the probability distribution learned by the NADE. $N=100$, $\beta=1$, and $N_H=64$.
}
\label{fig:HistoBeta1}
\end{center}
\end{figure}

\begin{figure}[t]
\begin{center}
\includegraphics[width=\linewidth]{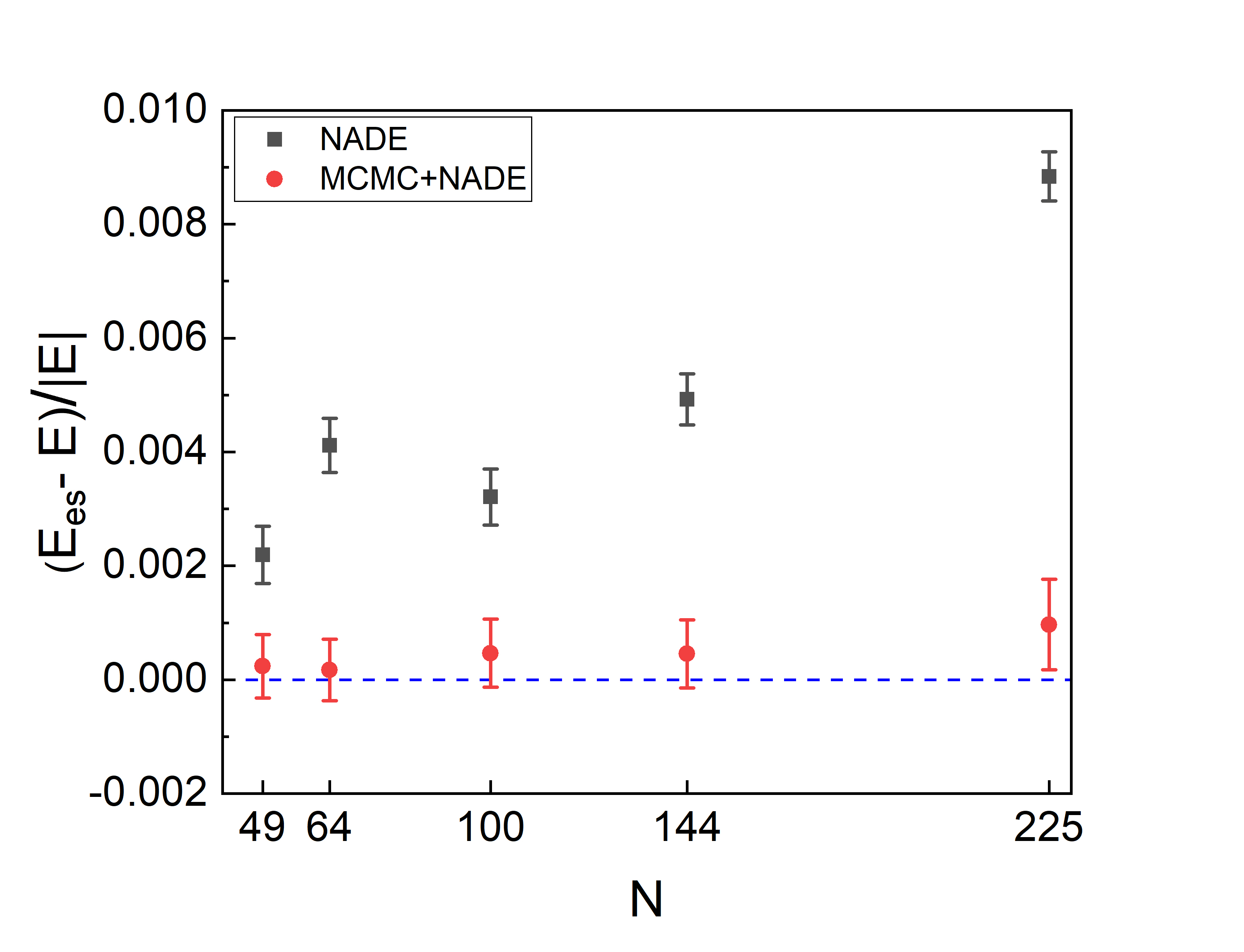}
\caption{Relative error $(E_{\mathrm{es}}-E)/|E|$ of the estimated energy $E_{\mathrm{es}}$ with respect to the exact value $E$, as a function of the system size $N$. For the red dots $E_{\mathrm{es}}$ is the expectation value over the probability distribution learned by NADE, while for the black squares $E_{\mathrm{es}}$ is the result of a NADE-driven MCMC simulation. $N_H=64$ and $\beta=1$.
}
\label{fig:Esize}
\end{center}
\end{figure}

Our main goal is to use a NADE to implement efficient MCMC simulations of the spin glass model~(\ref{H}), even in the low-temperature regime.
As a preliminary step, we show that the NADE can be trained to mimic the Boltzmann distribution in an unsupervised learning scheme.
As an illustrative example, the case of a spin glass of size $N=100$  at $\beta=1$ is analyzed in Fig.~\ref{fig:loglikelihoodH}. The total training set includes $10^5$ configurations, generated using single spin-flip updates. Every $800$-th configuration sampled by the Markov chain is included in the training set (corresponding to a total of $8\times 10^7$ single spin-flip MCMC steps). This allows suppressing the statistical correlations among the training configurations. In each training epoch, $2\times 10^4$ configurations are randomly selected.
The total training set is considered to be representative of the Boltzmann distribution, since at $\beta=1$ the single spin-flip algorithm is still sufficiently efficient to explore all physically relevant regions of the configuration space within the feasible simulation times. 
The training is performed by minimizing the negative log-likelihood of the training set via stochastic gradient descent, as explained in the previous section. The learning rate is $\eta=10^{-3}$ and the mini-batch size is $N_b=16$. The results obtained with reasonable variations around these values are comparable.
After $N_e\simeq 100$ learning epochs the training appears to have converged, slightly depending on the number of hidden units $N_H$.
To verify that the NADE is not overfitting the training data, and to quantify its generalization accuracy, we compare the energy expectation value $E_{\mathrm{NADE}}$ over the probability distribution learned by the NADE (see Eq.~\eqref{Enade}), with the exact result $E$. 
This comparison is shown in Fig.~\ref{fig:Enade}. For small $N_H$ a sizable bias occurs, indicating that a small NADE is not flexible enough to accurately reproduce the Boltzmann distribution. However, with $N_H\approx 100$ hidden neurons the bias is smaller than the statistical error-bars.
To perform a more stringent energy-resolved benchmarking, we compare the histograms of the energies sampled by the NADE probability distribution with the values sampled during the single spin-flip MCMC simulation (see Fig.~\ref{fig:HistoBeta1}). Excellent agreement is found in all relevant energy regimes.
This is a remarkable finding of our work, in view of the complex structure of the spin-glass configuration-space in this intermediate temperature regime.
As shown if Fig.~\ref{fig:Esize}, the accuracy seems to slowly decrease with the system size $N$, if $N_H$ is fixed.

\begin{figure}[b]
\begin{center}
\includegraphics[width=\linewidth]{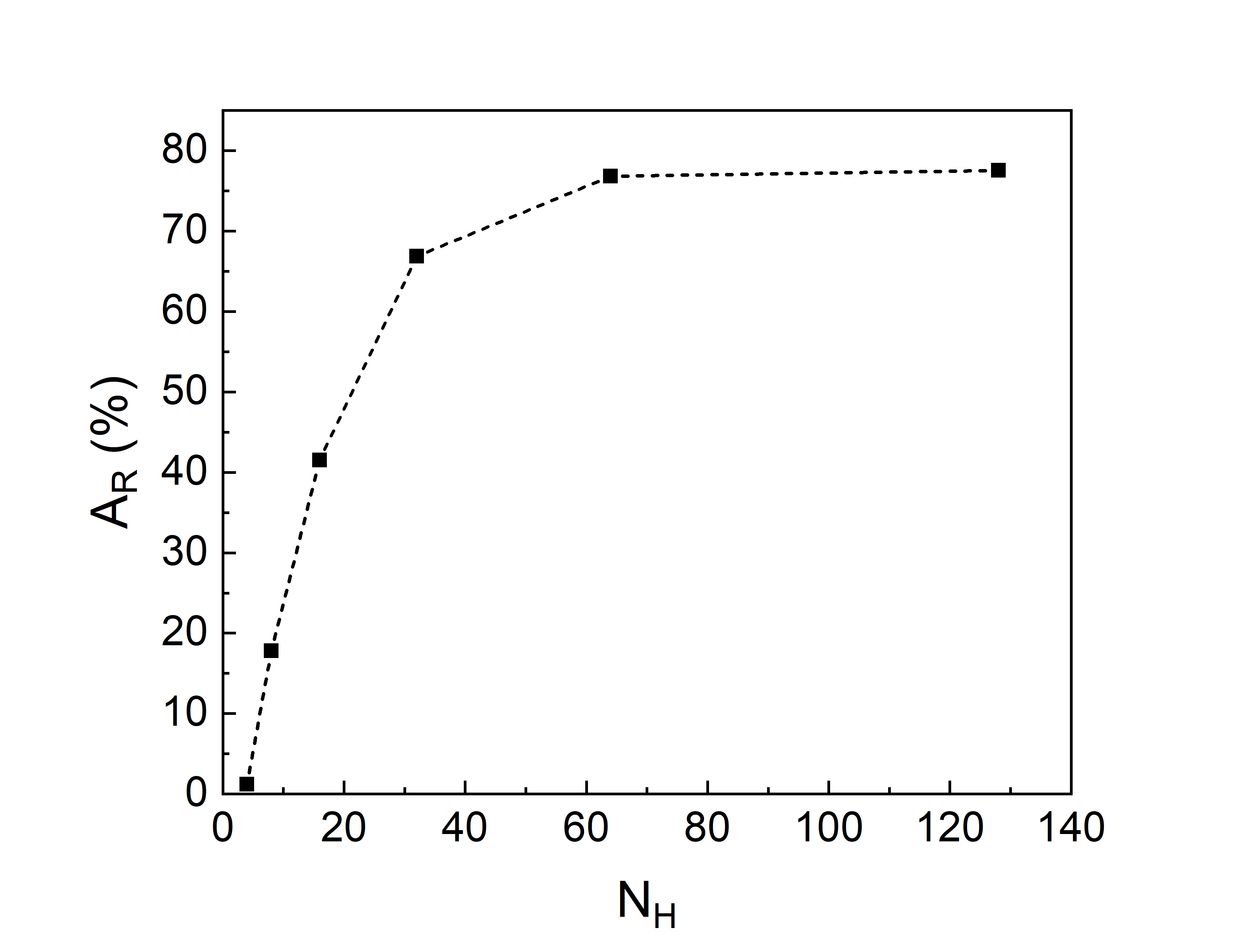}
\caption{Percentage acceptance ratio $A_R$ in NADE-driven MCMC simulations, as a function of hidden-neuron number $N_H$. $N=100$ and $\beta=1$.}
\label{fig:RH}
\end{center}
\end{figure}

\begin{figure}[t]
\begin{center}
\includegraphics[width=\linewidth]{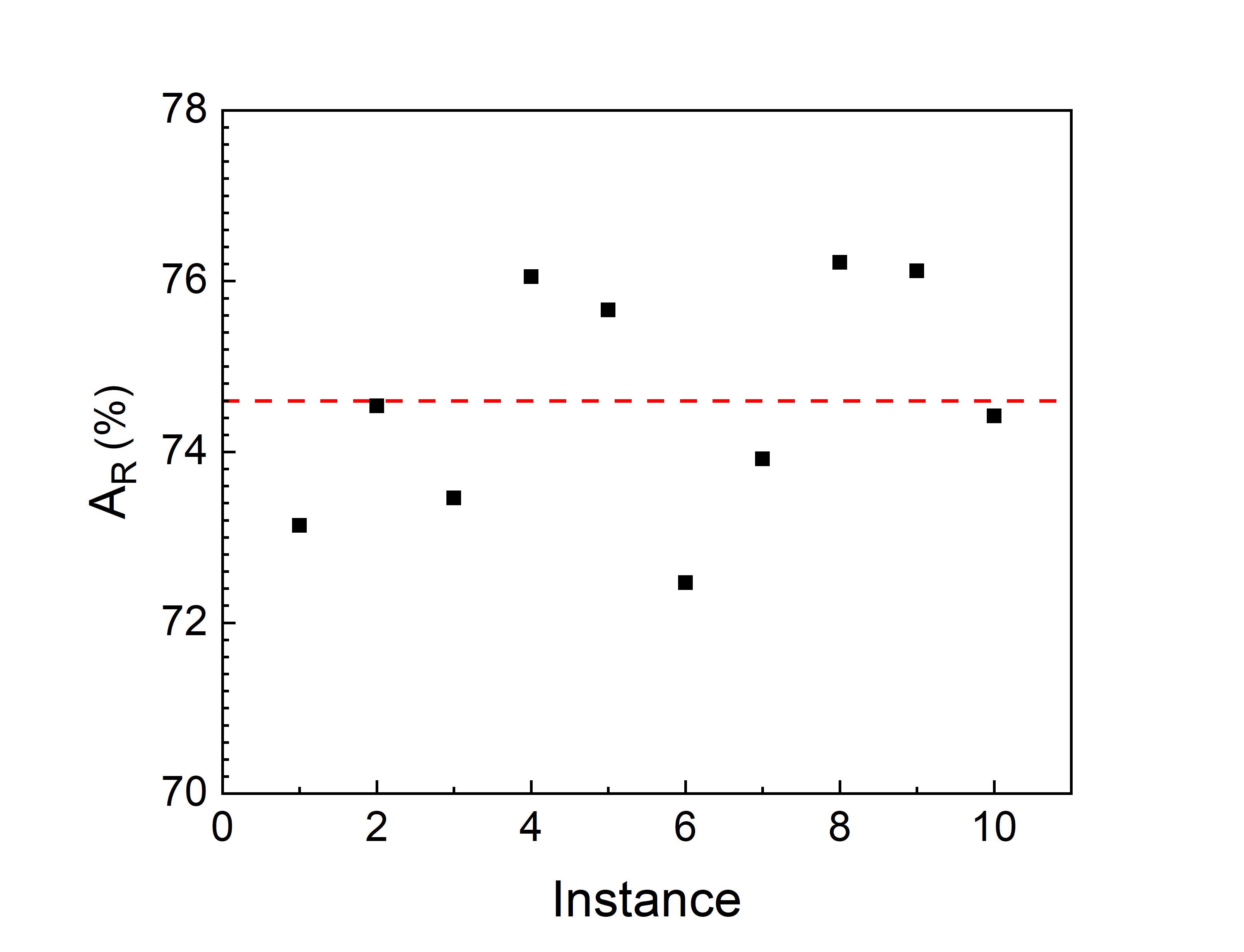}
\caption{Percentage acceptance ratio $A_R$ in NADE-driven MCMC simulations for 10 instances of the spin-glass Hamiltonian with different realizations of the gaussian random couplings. The red dashed line indicates the average value. $N=100$, $\beta=1$, and $N_H=64$.
}
\label{fig:RR}
\end{center}
\end{figure}

\begin{figure}[b]
\begin{center}
\includegraphics[width=\linewidth]{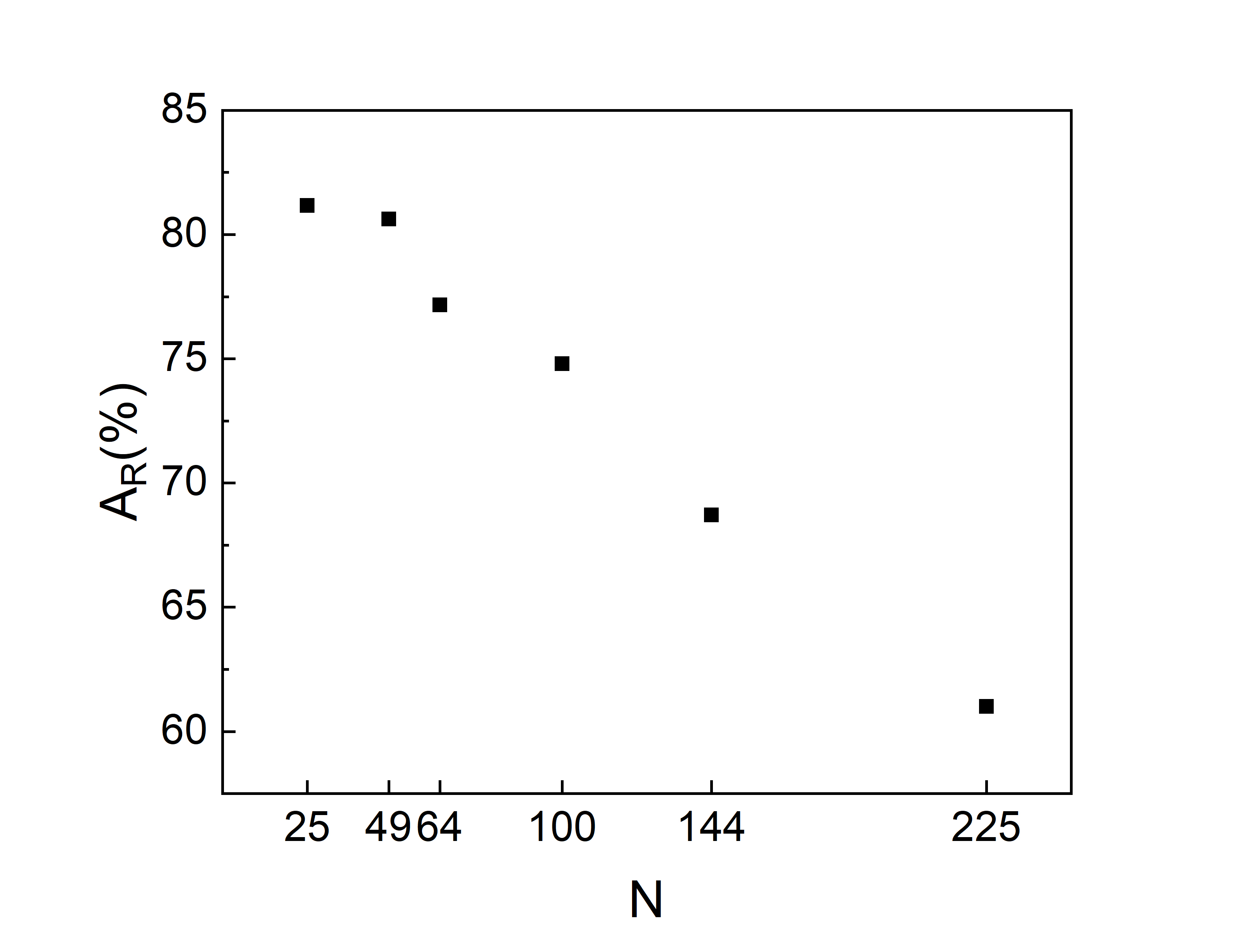}
\caption{Percentage acceptance ratio $A_R$ in NADE-driven MCMC simulations as a function of the system size $N$.  $\beta=1$ and $N_H=64$.
}
\label{fig:RN}
\end{center}
\end{figure}

By combining the trained NADE with the Metropolis-Hastings algorithm one can, on the one hand, remove the residual bias in the NADE expectation value, and, on the other hand, boost the efficiency of the MCMC simulation.
The trained NADE is used to define the proposal distribution for the  Metropolis-Hastings algorithm. As explained in the previous section, this leads to an efficient algorithm if the acceptance rate $A_R$, namely the percentage of accepted updates, is high. We recall that if the probability distribution learned by NADE exactly matches the Boltzmann distribution, the acceptance rate is $A_R=100\%$. As shown in Fig.~\ref{fig:RH}, for $N_H=10$ one has $R\simeq 20\%$, but for $N_H\simeq 100$ the acceptance rate is as high as $R\simeq 80\%$. $A_R$ appears to saturate for large $N_H$, possibly indicating that a larger dataset or a more powerful optimization algorithm are needed to further train large NADEs.
Regularization methods might also improve the results.
To verify that for different realizations of the gaussian couplings one obtains comparable results, we show in Fig.~\ref{fig:RR} the acceptance ratios for $10$ instances of the spin-glass model (\ref{H}). The relative variations are smaller than $10\%$, indicating that the performance of NADE does not strongly depend on the specific instance considered.
Instead, the acceptance rate slightly decreases with the system size $N$ if the number of hidden neurons $N_H=64$ is kept fixed (see Fig.~\ref{fig:RN}), reaching $R\simeq 60\%$ for $N=225$. This suggests that more hidden neurons or larger training sets might be required when the system size increases.
Importantly, the predictions provided by the NADE-driven MCMC simulations always agree with the exact results for the corresponding systems size. The latter value is computed via long single spin-flip MCMC simulations, including $8\times 10^7$ MCMC steps, verifying that simulations started from different configurations agree within the statistical uncertainties.The latter are determined by the standard blocking method~\cite{newman1999monte}. 
This agreement confirms that the NADE probability distribution has a sizable weight in all physically relevant regions of the configuration space, allowing an ergodic MCMC simulation.
For small $N_H$ one obtains larger error-bars due to the lower acceptance rate (see Fig.~\ref{fig:Enade}), which leads to longer correlation times.
The NADE-driven MCMC simulation is unbiased even for the largest system size addressed in this work (see Fig.~\ref{fig:Esize}), for which the NADE less accurately mimics the Boltzmann distribution.
The histogram of the configuration energies sampled during the NADE-driven MCMC simulation agrees with the one corresponding to a long single-spin flip simulation (see Fig.~\ref{fig:HistoBeta1}), confirming that the probability distribution learned by NADE has essentially the same support as the Boltzmann distribution.

\begin{figure}[t]
\begin{center}
\includegraphics[width=\linewidth]{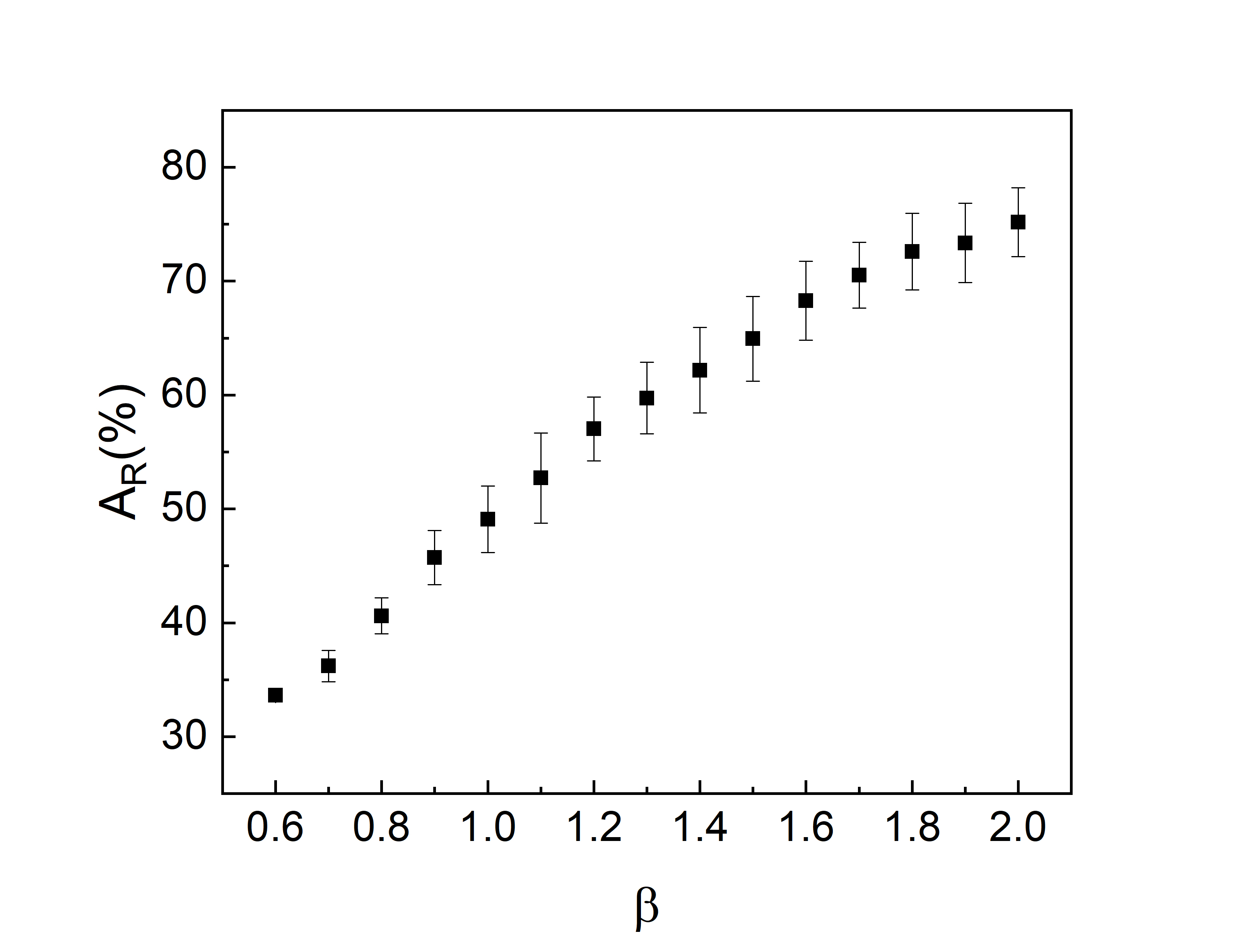}
\caption{Percentage acceptance ratio $A_R$ in the NADE-driven MCMC simulations of the sequential tempering procedure, as a function of the inverse temperature $\beta$. $N=100$ and $N_H=64$.
}
\label{fig:ARseq}
\end{center}
\end{figure}

\begin{figure}[b]
\begin{center}
\includegraphics[width=\linewidth]{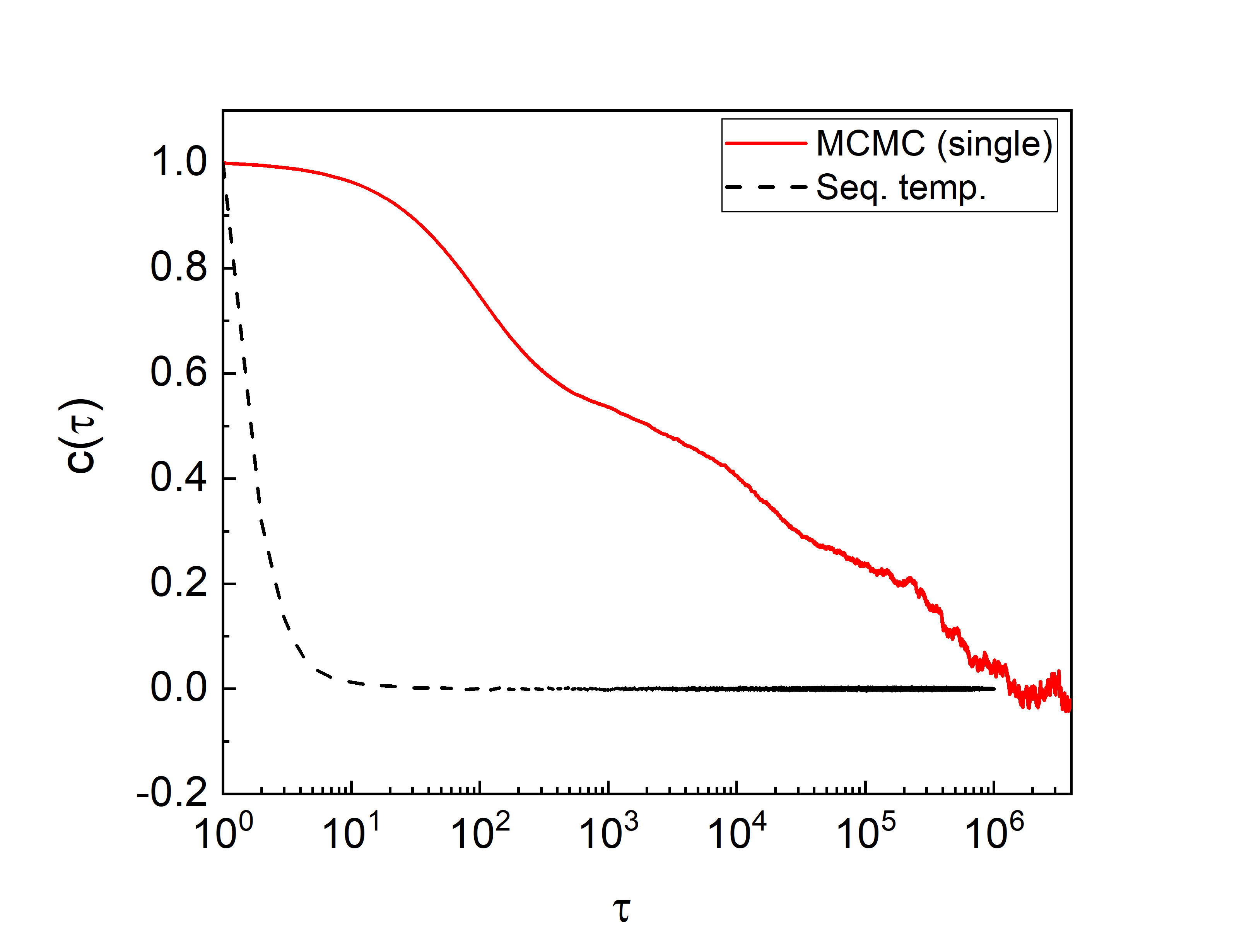}
\caption{Autocorrelation function $c(\tau)$ of the configuration energies at $\beta=2$. The red line corresponds to the single spin-flip MCMC algorithm (data averaged of $25$ simulations started from different initial configurations), while the black dashed line corresponds to the last NADE-driven MCMC simulation of the sequential tempering procedure. 
$N=100$ and $N_H=64$. 
}
\label{fig:ACF2}
\end{center}
\end{figure}

\begin{figure}[t]
\begin{center}
\includegraphics[width=\linewidth]{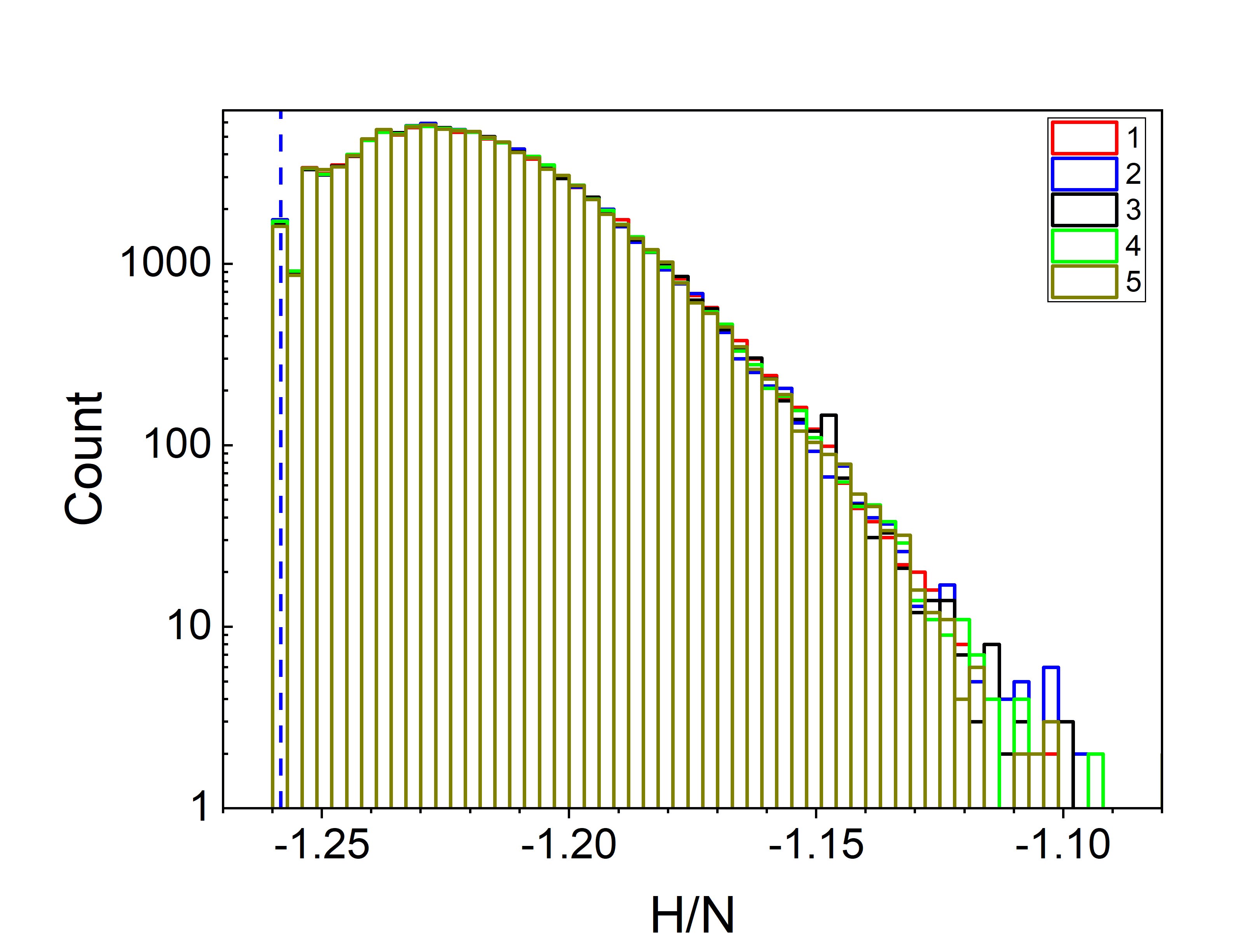}
\caption{Histogram of $10^5$ configuration energies per spin $H/N$ generated in the last NADE-driven MCMC simulation  of the sequential tempering procedure at $\beta_{\mathrm{max}}=2$. Every $10$-th configuration is counted. 
The 5 datasets correspond to just as many sequences started from different initial configurations at $\beta_0=0.5$, for the same realization of the spin-glass model. $N=100$ and $N_H=64$.
The blue vertical dashed line indicates the ground-state energy for this instance of the spin-glass model,  obtained from the spin-glass server~\cite{spinglassserver}.
}
\label{fig:ACF5}
\end{center}
\end{figure}

\begin{figure}[b]
\begin{center}
\includegraphics[width=\linewidth]{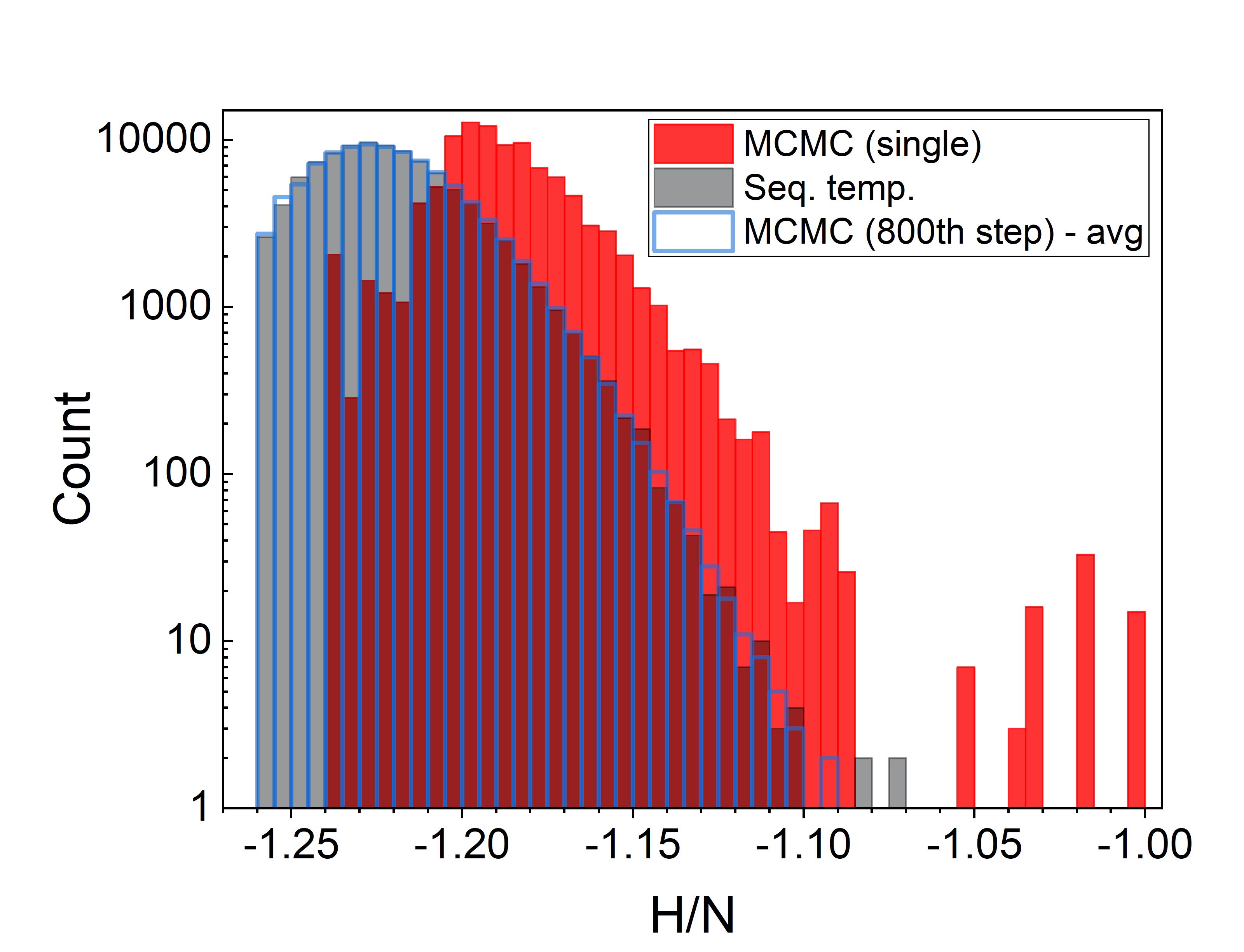}
\caption{Histogram of $10^5$ configuration energies per spin $H/N$ sampled at $\beta=2$. 
The empty blue columns correspond to the single spin-flip MCMC algorithm (data averaged over $25$ simulations started from different configurations, ran for $8\times 10^7$ steps, every $800$-th configuration is counted). The gray columns correspond to the last NADE-driven MCMC simulation of the sequential tempering procedure (every $10$-th configuration is counted). 
For comparison, the full red columns indicate the first $10^5$ configurations sampled via the single spin-flip MCMC algorithm.
The vertical dashed line indicates the ground-state energy, obtained from the spin-glass server~\cite{spinglassserver}.
$N=100$ and $N_H=64$.
}
\label{fig:Histobeta2}
\end{center}
\end{figure}

\begin{figure}[t]
\begin{center}
\includegraphics[width=\linewidth]{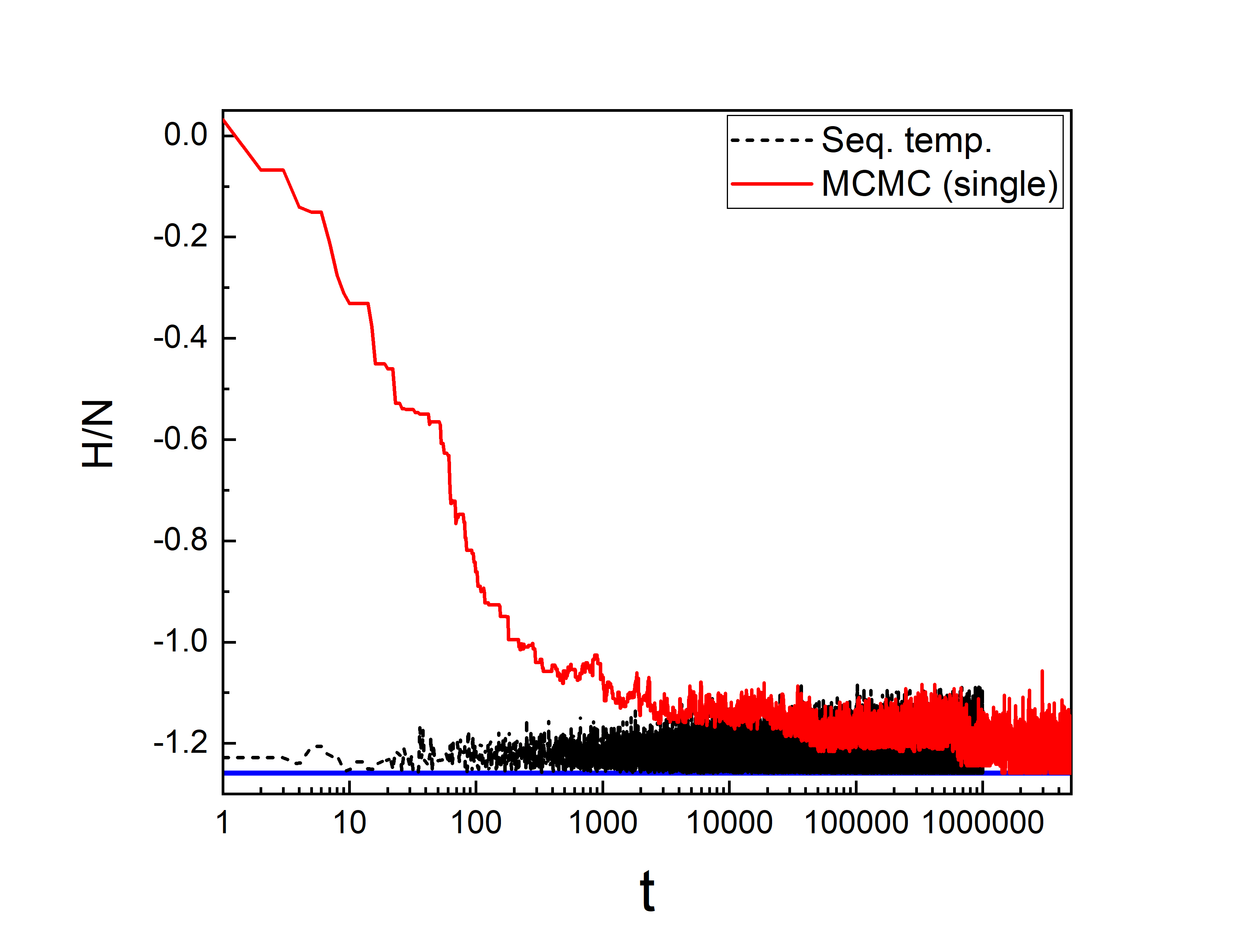}
\caption{Configuration energy per spin $H/N$ as a function of the number of MCMC steps $t$, at $\beta=2$. Dashed black line corresponds to the last NADE-driven MCMC simulation of the sequential tempering procedure. Solid red line corresponds to the single spin-flip MCMC simulation. The horizontal blue line indicates the ground-state energy, obtained from the spin-glass server~\cite{spinglassserver}. $N=100$ and $N_H=64$.
}
\label{fig:Hbeta2}
\end{center}
\end{figure}

The unsupervised learning scheme described in the previous paragraph assumes that a training dataset representative of the Boltzmann distribution can be generated. 
As explained above, with the single spin-flip algorithm this becomes unfeasible for inverse temperatures $\beta \gg 1$ due to the diverging correlation times. Clearly, one could adopt more sophisticated global-update MCMC algorithms as, e.g., the isoenergetic cluster updates~\cite{houdayer2001cluster,zhu2015efficient}, and then use the NADE only to accelerate the computation of physical properties in a second MCMC simulation. 
As we discuss in the following, a simple procedure can be implemented to efficiently simulate the low temperature regime, even without employing global-update methods to generate the training dataset. We dub this procedure sequential tempering, in analogy with the popular parallel tempering method~\cite{hukushima1996exchange}.
This procedure begins by training a NADE at a relatively low inverse temperature $\beta_0$, where ergodic MCMC simulations are feasible even with single spin-flip algorithms.
Next, various NADE-driven MCMC simulations are run at the inverse temperatures $\beta_s=\beta_{s-1}+\delta\beta$, where $\delta\beta$ is a small increment and $s=1,2,\dots,s_{\mathrm{max}}$, using the NADE trained at $\beta_{s-1}$ as proposal distribution.
In our numerical experiment with an Ising glass of size $N=100$ and a NADE with $N_H=64$ hidden neurons, we choose $\beta_0=0.5$, $\delta\beta=0.1$, and $s_{\mathrm{max}}=15$, allowing us to reach a low temperature corresponding to $\beta_{\mathrm{max}}=2$. 
We emphasize that at this final temperature even $8\times 10^7$ single spin-flip MCMC steps might not be sufficient to guarantee ergodicity. 
In each NADE-driven MCMC simulation, every $10$-th configuration is stored to train the next NADE. As discussed below, this is sufficient to suppress the residual statistical correlations. The total training-set size is again $N_t=10^5$.
Since each NADE is employed as proposal distributions at a different temperature compared to the training temperature, one should expect a decreased acceptance rate $A_R$. 
As shown in Fig.~\ref{fig:ARseq}, in the first few steps of the sequence the acceptance rate is indeed moderately low, $R \approx 40\% $. However, it increases up to  $R \approx 75\% $ in the low-temperature regime. This indicates that the NADE is able to closely follow the evolution of the Boltzmann distribution as the temperature decreases, identifying the regions of configuration space where the weight is sizable. 
Notably,  each subsequent training stage can be significantly accelerated by initializing the NADE parameters to the final values of the previous learning stage.
It is worth emphasizing again that with different disorder realizations the NADEs provide comparable results. Indeed, the data shown in Fig.~\ref{fig:ARseq} correspond to the average of $5$ instances of the random couplings.
The NADE-driven MCMC simulations are ergodic and efficient even in the low-temperature regime. The correlation function at $\beta=2$ displays a sharp drop (see Fig.~\ref{fig:ACF2}), confirming that the correlation times are minimal, as anticipated above. Instead, with the single spin-flip algorithm even after $t\sim 10^6$ MCMC steps the statistical correlations are sizable. 
This important finding indicates that the sequential tempering allows us to efficiently sample the low-temperature Boltzmann distribution.
This statement is confirmed by the precise agreement of the $5$ histograms shown in Fig.~\ref{fig:ACF5}, which correspond to just as many sequences for the same instance of the random couplings started from different configurations at $\beta_0=0.5$.
In order to obtain a comparable histogram with the single spin-flip algorithm, we have to perform as many as $8\times 10^7$ steps, and to average over $25$ runs started from different configurations (every $800$-th sampled energy is counted). The comparison with the sequential-tempering histogram is shown in Fig.~\ref{fig:Histobeta2}. For illustrative purposes, Fig.~\ref{fig:Histobeta2} displays also the histogram of the first $10^5$ energies sampled in the initial portion of a single spin-flip MCMC run. Clearly, this dataset is highly biased, in particular in the low-energy regime where the weight is negligible. This indicates that the single spin-flip algorithm requires many more steps to reach low-energy configurations.
To shed more light on the equilibration dynamics, it is useful to visualize how the configuration energy $H(t)$ evolves along the last NADE-driven MCMC simulation of the sequential tempering (see Fig.~\ref{fig:Hbeta2}). The equilibration time appears to be negligible. In particular, this simulation touches a ground-state configuration after as few as $t=31$ MCMC steps. This ground-state energy is obtained from the spin-glass server~\cite{spinglassserver} at the  University of Cologne, which implements an exact polynomial-time algorithm for two-dimensional lattices.
This phenomenology appears to be general: repeating $5$ sequential tempering procedures for the same instance of the random couplings but different initial configurations at $\beta_0$, or for $5$ different instances of the random couplings, we always find that the ground-state energy is sampled within $t \lesssim \sim 10^2$ MCMC steps.
For comparison, the configuration energies obtained from a single spin-flip MCMC simulation are also shown in Fig.~\ref{fig:Hbeta2}. In this specific case,  a ground-state configuration is reached only after $t\sim 10^6$ MCMC steps. 
In fact, an (admittedly incomplete) analysis shows that with $t \sim 10^7$ single spin-flip steps only $\sim 50\%$ of the times the ground-state energy is sampled.
This illustrates the well-known difficulty in identifying ground-state configurations via  single spin-flip Metropolis-Hastings updates.

The findings discussed above indicate that the sequential tempering procedure allows one to efficiently simulate the low-temperature regime of a short-range spin-glass model. 
Furthermore, they suggest that NADEs could be employed to boost the efficiency of stochastic optimization methods for binary optimization problems.
Indeed, heuristic optimization methods like simulated annealing~\cite{kirkpatrick1983optimization} exploit MCMC algorithms to explore the solution space.
Generic binary optimization problems can be mapped to disordered Ising Hamiltonians analogous to Eq.~\eqref{H}. The long correlation times discussed above also occur when tackling the optimization problem, often preventing the optimal solution from being found. 
Our results point to the use of NADEs trained with a sequential tempering procedure as the engine for simulated-annealing optimizations.
It is also worth mentioning that NADEs could also be trained from data produced using physical quantum annealers, such as the devices commercialized by D-Wave Systems (see, e.g, Ref.~\cite{boixo2014evidence}). These devices are special-purpose adiabatic quantum computers designed to solve quadratic unconstrained optimization problems. They allow sampling low-energy configurations of programmable Ising  Hamiltonians, with the goal to identify the optimal solution.
The trained NADE could be employed to drive simulated-annealing optimizations. This might help eliminating the effect of the inevitable noise present in the values of the model parameters programmed on the physical device.

\section{Conclusions}
\label{conclusions}
We have shown how to use an autoregressive generative neural network, namely a NADE~\cite{larochelle2011neural}, to boost the efficiency of Markov chain Monte Carlo (MCMC) simulations of a two-dimensional Ising Hamiltonian with nearest-neighbor gaussian couplings. This model Hamiltonian is an archetype of spin-glass theory.
The NADEs have been trained within an unsupervised learning scheme, which consists of minimizing the Kullback-Leibler divergence with respect to a dataset of configurations generated via standard MCMC simulations.
Our analysis quantified how accurately a NADE can mimic the Boltzmann distribution of a spin-glass model, depending on the number of hidden neurons and the number of visible spins.
The trained NADEs have then been used as proposal distributions in smart Monte Carlo simulations based on the Metropolis-Hastings algorithm. 
This allowed us to implement efficient global updates, whose computational cost is linear in the system size.
In particular, we have implemented a sequential tempering procedure. Starting from a higher temperature, the procedure reaches the low temperature regime via a sequence of MCMC simulations and training stages performed at successively lower temperatures. This allowed us to run efficient MCMC simulations with very short autocorrelation times, even in regimes where computing thermodynamic properties with standard local algorithms is difficult, if not totally unfeasible.
Furthermore, it has been verified that at low temperatures the NADE-driven MCMC simulations quickly sample ground-state configurations. This result suggests to employ autoregressive neural networks in combination with simulated annealing or other stochastic methods to solve binary optimization problems.

Our work complements other recent investigations on the use of autoregressive models for classical mechanics problems~\cite{wu2019solving,nicoli2019asymptotically}. We described the use of an unsupervised learning scheme, instead of reinforcement learning, and we tackled a short-range spin-glass Hamiltonian instead of clean systems or mean-field infinite-range disordered models. 
It is worth stressing that the unsupervised learning scheme could be combined with any of the sophisticated MCMC techniques that have been developed over the years to simulate spin glasses. Specifically, the global-update methods could be employed to efficiently generate training datasets. The NADE could then be used to speed up the computation of physical properties, including observables that explicitly depend on the partition function~\cite{nicoli2019asymptotically}.
The results we have presented in this article are encouraging, and raise ambition to further investigations, including different spin-glass models or deeper generative neural networks such as PixelCNN~\cite{oord2016pixel}, variational autoencoder~\cite{kingma2013auto}, and generative adversarial networks~\cite{goodfellow2014generative}. It would also be important to analyze observables other than the average energy as, e.g., the spin-spin correlations or the Edwards-Anderson order parameter. We leave these endeavors to future studies.

\section*{Acknowledgements}
\noindent 
The authors thank I. Murray and G. Carleo for useful discussions. Financial support from the FAR2018 project titled ``Supervised machine learning for quantum matter and computational docking'' of the University of Camerino and from the Italian MIUR under the project PRIN2017 CEnTraL 20172H2SC4 is gratefully acknowledged. S. P. also acknowledges the CINECA award under the ISCRA initiative, for the availability of high performance computing resources and support. 
\bibliography{Ref}

\end{document}